# Analysis of Competence Level and the Attendance of the Lecturer in Its Effects on Students Grade Using Fuzzy Quantification Theory

**Hindayati Mustafidah[1], Suwarsito[2]**

[1] Informatics Engineering, Muhammadiyah University of Purwokerto
Purwokerto, Central Java, Indonesia, 53182

[2] Geographic Education, Muhammadiyah University of Purwokerto
Purwokerto, Central Java, Indonesia, 53182

**Abstract**
It known that teachers as educators should have competence that shows its quality so that the lecture material provided can be absorbed by the student. The competencies in this term include the pedagogic, professional, personality, and social. The competence that owned by lecturer can be obtained from the results of the assessment conducted by the student through the filling of the questionnaire. This study conducted an analysis of the level of the lecturer competence of the relationship between the present of lecturer in classroom with a percentage of the value of passing students in courses using fuzzy quantification theory. Based on the results of the four competencies acquired professional competence that contributes most of 79.45% in contributed the attendance of lecturer will it affect the percentage of passing students in courses that are shown with a percentage of the graduation minimum B.

## 1. Introduction

Some of the influence factors of student learning achievement, namely: (1) teachers/lecturers, (2) students, and (3) infrastructure. Teacher/lecturer factors include the ability to teach, the mastery of the material and its quality. The purpose of teaching and learning in a College can be reached from the lecturer and students factors. The liveliness of the lecturers in giving lessons and liveliness condition of the student in following lessons is a major key for the success of the process of teaching and learning. The success of the process of teaching and learning, for students, it can be seen with the measuring instrument in the form of final grade obtained. Usually, a student is said to have good grade in a subject, when the students get a grade greater than or equal to ' B ' (grade ≥ B). Similarly, a lecturer said to be a success in the process of teaching and learning, when the value of the performance is also good. Therefore, in the study conducted an analysis of the influence of connection performance/competence of the lecturers with liveliness gave a lecture to the achievement of student learning (grade ≥ B).

The competence of a lecturer has a corresponding to his discipline that based on the knowledge, experiences, skills, creativity, initiative, motivation as a lecturer and a positive work culture, which in the end have expertise in accordance with the needs of a lecturer at a college [1]. The competence of a lecturer can eventually provide services at various parties who need it, especially to the students who constantly interact on him. The competence of a lecturer who is able to work and always ready to conform to the standards of service that are required by the student will be able to contribute to the achievement of the vision and mission of a college. Other accounts state that competency is the ability, competence and skills that a person with regard to the duties of the position or profession [2]. Lecturers are professors at the College, which has the main task of teaching as his profession. Thus, according to Rumanti in [3], competence of lecturer is a communication skills, managerial or leadership ability, the ability to get along or build relationships, personalities are intact or honest, as well as the rich idea and creative to be belonged by a lecturer in order to run their practices. In [4] a college professor is said to be competent when he has mastered the four basic competences, namely pedagogic, professional, personality, and social. In a study conducted by [5] revealed that there is a correlation of pedagogic competencies influence the learning achievements of students. In this study are discussed regarding the types of competence that have a connection with the lecturer attendance rates that have the greatest contribution to the graduation rate of students in courses. This competence was obtained from assessment that performed by the student. The assessment conducted through questionnaire is qualitative assessment scale from not very good until very good with a score range of 1 – 5. In this research used Fuzzy Quantification Theory (FQT). FQT is known as a method to control the qualitative data by using fuzzy set theory. Control is intended more to explain the fuzzy events using the values in the range [0, 1] that express a qualitative opinions [6]. The aim of Fuzzy Quantification Theory (qualitative regression analysis) is determines the relationship between qualitative variables are given a value between 0 to 1, and the numerical variables in a fuzzy group given in the sample.

## 2. Method

The collection of data in this study use questionnaire (to get the data of competence level of the lecturers) and documentation (to obtain data the level of attendance lecturer in giving lectures and student graduation rate in the courses given by the lecturer). Rate of the lecturer's attendance give lectures data are obtained from Quality Assurance in Muhammadiyah University of Purwokerto. The data are attendance lecturer every course held at the Muhammadiyah University of Purwokerto at academic year 2011/2012. While the students achievement data obtained from the Academic Bureau of Muhammadiyah University of Purwokerto for academic year 2011/2012.

The operational steps undertaken in this study are determines external data, determines the data category, and building fuzzy group. The final step is applying fuzzy quantification theory and analyzing each factor of the level of competence of the lecturers that affect the student level of graduation in course.

## 3. Results and Discussion

### 3.1 External Data

Lecturer's competence data are obtained from questionnaire that filled by students. Sample have been taken from several students from various faculties, namely the Faculty of Engineering, Education, Healthy, Agriculture, Psychology, Literature, and Pharmacy that amounted to 433 data records.

### 3.2 Category and Fuzzy Group

The variables in this study consist of the competence level of lecturer, attendance lecturer in giving lectures, and student graduation rate in enrolling lectures, indicated by percentage of the minimum grade of B (grade ≥ B) in each of the courses. The category is used only one level of attendance lecturer in classrooms. Lecturer attendance is said to be HIGH by following the membership function on chart patterns like in Fig. 1. While the fuzzy group in the form of competence i.e. 1) pedagogic 2) professional, 3) personality, and 4) social.

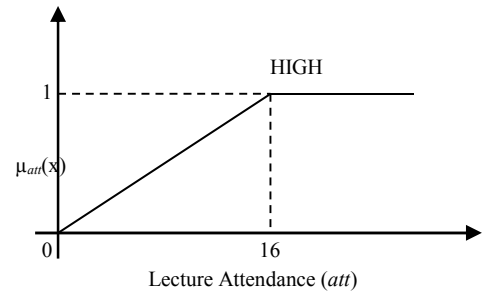

Fig. 1 Fuzzy set of lecturer's attendance in classroom.

The membership function of the level of lecturer's attendance is presented in Eq. 1 as follows.

$$\mu_{att}(x) = \begin{cases} \frac{x}{16}; & 0 \leq x \leq 16 \\ 1; & > 16 \end{cases} \quad (1)$$

Using linear regression obtained relationship between the level of lecturer attendance (x) with percentage of grade ≥ B obtained a student regardless of other attributes like in Eq. 2 and 3 as follows.

$$y = 0.428x + 62.953 \quad (2)$$

or

$$y = 8.727\mu(x) + 61.498 \quad (3)$$

### 3.3 Factor Analysis (Fuzzy Group)

The variables in this study consist of the competence level of lecturer, attendance lecturer in giving lectures, and student graduation rate in enrolling lectures, indicated by percentage of the minimum grade of B (grade ≥ B) in each of the courses. The category is used only 1 level of attendance lecturer in classrooms. Lecturer attendance is said to be HIGH by following the membership function on chart patterns like in Fig. 1. While the fuzzy group in the form of competence i.e. 1) pedagogic 2) professional, 3) personality, and 4) social.

**Fuzzy group analysis 1: pedagogic competence.** The relationship between categories and associated lecturer in attendance rates HIGH and competence of pedagogic to earn a percentage grade ≥ B indicated by characteristics of the FQT as shown in the following Table 1.

Table 1: The characteristics of FQT for pedagogic competence fuzzy group

| Sample | Fuzzy Group B $\mu_B(k)$ | x | y |
|---|---|---|---|
| | Pedagogic Competence | Lecturer's Attendance | Percentage of grade ≥ B |
| 1 | 0.938 | 0.813 | 50 |
| 2 | 0.870 | 0.813 | 66.67 |
| 3 | 0.870 | 0.750 | 81.82 |
| 4 | 0.958 | 0.750 | 88.89 |
| 5 | 0.938 | 0.813 | 30.77 |
| 6 | 0.965 | 0.750 | 58.33 |
| 7 | 0.938 | 0.938 | 81.82 |
| 8 | 0.870 | 0.750 | 80 |
| 9 | 0.870 | 0.813 | 78.57 |
| 10 | 0.958 | 0.813 | 100 |
| … | | | |
| 433 | 0.868 | 0.813 | 72.06 |

Subsequently formed matrix y which is a matrix vector of percentages grade ≥ B, while the matrix G is a diagonal matrix rectilinear sized 433x433 with its elements contain $\mu_B(k)$ that is the value of the membership of the sample (in this case k = 1) on fuzzy group B (pedagogic competence) and other elements of zero (0). Matrix X sized 433x1 with the row element of $k^{th}$ is $\mu_1(k)$, namely the degree of membership of the sample on a category level of lecturer's attendance in HIGH. Elements of the matrix X and y for the fourth fuzzy group is the same, only on a different element of the matrix G.

$$G = \begin{vmatrix} 0,938 & 0 & 0 & 0 & 0 & 0 & 0 & 0 \\ 0 & 0,870 & 0 & 0 & 0 & 0 & 0 & 0 \\ 0 & 0 & 0,870 & 0 & 0 & 0 & 0 & 0 \\ 0 & 0 & 0 & 0,958 & 0 & 0 & 0 & 0 \\ 0 & 0 & 0 & 0 & 0,938 & 0 & 0 & 0 \\ 0 & 0 & 0 & 0 & 0 & 0,965 & 0 & 0 \\ \dots & \dots & \dots & \dots & \dots & \dots & \dots & \dots \\ 0 & 0 & 0 & 0 & 0 & 0 & 0 & 0,868 \end{vmatrix}$$

$$X = \begin{vmatrix} 0.813 \\ 0.813 \\ 0.750 \\ 0.750 \\ 0.813 \\ \dots \\ 0.813 \end{vmatrix} \quad y = \begin{vmatrix} 50 \\ 66.67 \\ 81.82 \\ 88.89 \\ 30.77 \\ \dots \\ 72.06 \end{vmatrix}$$

Determination of category a are minimizing the error variance is given by the following Eq. 4.

$$a = (X'GX)^{-1}X'Gy \qquad (4)$$

Based on the obtained Eq. 4, a = 87.9233, so y = 87.9233 μ(x) or y = 5.4823x. Based on the calculations obtained, seem that taking into the level of pedagogic competencies will contribute an additional 5.0543 (5.4823 – 0.428) toward the acquisition of the percentage grade ≥ B against classes, or of 79.1963% (87.9233% - 8.727%). The intersection point of the two lines of regression occurs at (0.7765, 63.2854), which means that the pedagogic competence of lecturer will give influence on the percentage of the grade ≥ B in the attendance of the lecturers give lectures on 12.42442 (0.7765 x 16) or above 12 times. Conversely, if the attendance of the lecturer give lectures at less than 12 times, then the level of pedagogic competence will not exert influence to increased achievement percentage of grade ≥ B.

**Fuzzy group analysis 2: professional competence.** The relationship between categories and associated lecturer in attendance rates HIGH and professional competence to gain a percentage of the grade ≥ B indicated by characteristics of the FQT as shown in the following Table 2.

Table 2: The characteristics of FQT for professional competence fuzzy group

| Sample | Fuzzy Group B $\mu_B(k)$ | x | y |
|---|---|---|---|
| | Professional Competence | Lecturer's Attendance | Percentage of grade ≥ B |
| 1 | 0.940 | 0.813 | 50 |
| 2 | 0.912 | 0.813 | 66.67 |
| 3 | 0.912 | 0.750 | 81.82 |
| 4 | 0.934 | 0.750 | 88.89 |
| 5 | 0.940 | 0.813 | 30.77 |
| 6 | 0.956 | 0.750 | 58.33 |
| 7 | 0.940 | 0.938 | 81.82 |
| 8 | 0.912 | 0.750 | 80 |
| 9 | 0.912 | 0.813 | 78.57 |
| 10 | 0.934 | 0.813 | 100 |
| … | | | |
| 433 | 0,828 | 0.813 | 72.06 |

As in the first fuzzy group, matrix y is a matrix vector of percentages of grades ≥ B, while matrix X sized 433x1

with the row element of $k^{th}$ is $\mu_1(k)$, namely the degree of membership of the sample of $k^{th}$ on a category level of lecturer's attendance in HIGH. The matrix G is a diagonal matrix rectilinear sized 433x433 with its diagonal elements contain $\mu_B(k)$ that the value of the sample membership of $k^{th}$ sample on fuzzy group B (professional competences) and other elements of zero (0).

By using the Eq. 4 obtained a = 88.1718, so
- y = 88,1718μ(x) or y = 5,498x
- addition of the contribution = 5.0700 (5,498 – 0.428) or 79.4448% (88,1718% - 8.727%).
- the intersection point of the two lines of regression occurs at (0.7741, 63.2843)
- analyzes: professional competence will exert influence on the percentage of the grade ≥ B in the attendance of the lecturer give lectures on 12.38556 (0.7741 x 4) or above 12 times. Conversely, if the attendance of the lecturer give lectures at less than 12 times, then the professional competence of the lecturers would not exert influence to increased achievement percentage of grade ≥ B.

**Fuzzy group analysis 3: personality competence.** The relationship between categories and associated lecturer in attendance rates HIGH and personality competence to gain a percentage of grade ≥ B indicated by characteristics of the FQT as shown in the following Table 3.

Table 3: The characteristics of FQT for personality competence fuzzy group

| Sample | Fuzzy Group B $\mu_B(k)$ Personality Competence | x Lecturer's Attendance | y Percentage of grade ≥ B |
|---|---|---|---|
| 1 | 0.955 | 0.813 | 50 |
| 2 | 0.931 | 0.813 | 66.67 |
| 3 | 0.931 | 0.750 | 81.82 |
| 4 | 0.942 | 0.750 | 88.89 |
| 5 | 0.955 | 0.813 | 30.77 |
| 6 | 0.961 | 0.750 | 58.33 |
| 7 | 0.955 | 0.938 | 81.82 |
| 8 | 0.931 | 0.750 | 80 |
| 9 | 0.931 | 0.813 | 78.57 |
| 10 | 0.942 | 0.813 | 100 |
| … | | | |
| 433 | 0,846 | 0.813 | 72.06 |

By using the Eq. 4 obtained a = 88.0678, so:
- y = 88.0678μ(x) or y = 5.4917x
- addition of the contribution = 5.0637 (5.4917 – 0.428) or 79.3408% (88.0678% - 8.727%).
- the intersection point of the two lines of regression occurs at (0.7751, 63.2847).
- analyzes: personality competence will exert influence on the percentage of the grade ≥ B in the attendance of the lecturer give lectures on 12.40179 (0.7751 x 16) or above 12 times. Conversely, if the attendance of the lecturer give lectures at less than 12 times, then the professional competence of the lecturers would not exert influence to increased achievement percentage of grade ≥ B.

**Fuzzy group analysis 4: social competence.** The relationship between categories and associated lecturer in attendance rates HIGH and social competence to gain a percentage of the grade ≥ B indicated by characteristics of the FQT as shown in the following Table 4.

Table 4: The characteristics of FQT for social competence fuzzy group

| Sample | Fuzzy Group B $\mu_B(k)$ Social Competence | x Lecturer's Attendance | y Percentage of grade ≥ B |
|---|---|---|---|
| 1 | 0.928 | 0.813 | 50 |
| 2 | 0.865 | 0.813 | 66.67 |
| 3 | 0.865 | 0.750 | 81.82 |
| 4 | 0.941 | 0.750 | 88.89 |
| 5 | 0.928 | 0.813 | 30.77 |
| 6 | 0.950 | 0.750 | 58.33 |
| 7 | 0.928 | 0.938 | 81.82 |
| 8 | 0.865 | 0.750 | 80 |
| 9 | 0.865 | 0.813 | 78.57 |
| 10 | 0.941 | 0.813 | 100 |
| … | | | |
| 433 | 0,857 | 0.813 | 72.06 |

By using the Eq. 4 obtained a = 87.8794, so:
- y = 87.8794μ(x) or y = 5.4801x
- additional contribution = 5.0521 (5.4801 – 0.428) or 79.1524% (87.8794% - 8.727%).
- the intersection point of the two lines of regression occurs at (0.7770, 63.2855)
- analyzes: social competence will exert influence on the percentage of the grade ≥ B in the attendance of the lecturer in classroom on 12.43131 (0.7770 x 16) or above 12 times. Conversely, if the attendance of the lecturer give lectures at less than 12 times, then the professional competence of the lecturers would not exert influence to increased achievement percentage of grade ≥ B.

As a result of the analysis using FQT, retrieved the weights corresponds to the four categories of fuzzy group as presented in Table 5.

Table 5: Category weights summary

| Fuzzy Group (competences) | Category Weights | | Additional Contribution | |
| --- | --- | --- | --- | --- |
| | As a coefficient $\mu(x)$ | As a coefficient $x$ | As a coefficient $\mu(x)$ | As a coefficient $x$ |
| Pedagogic | 87.9233 | 5.4823 | 79.1963 | 5.0543 |
| Professional | 88.1718 | 5.498 | 79.4448 | 5.0700 |
| Personality | 88.0678 | 5.4917 | 79.3408 | 5.0637 |
| Social | 87.8794 | 5.4801 | 79.1524 | 5.0521 |

Based on Table 5 looks that the second fuzzy group namely professional competence contribute most in attendance rates accounted for a lecturer in obtained percentage grade of minimal B than other competencies, despite minimal requirements for the attendance of lecturer is the same among the four competencies i.e. 12 times. This means that of the 433 the data taken from the condition of a lecturer at the Muhammadiyah University of Purwokerto, there is some level of professional competence of the lecturers are still lacking. By having a high level of professional competence, the lecturer will increasingly have the ability to transfer knowledge to students. This also means that by having a professional level of competence that is triggered by the attendance of a lecturer in classroom, will impact positively on the grade achieved by the student.

## 4. Conclusions

Conclusion of this study is that by using fuzzy quantification theory, the professional competence of the lecturers contribute most in attendance of lecturer in obtained its effect on the percentage value of the graduation courses minimum of B, i.e. with the contribution of 79,45%. This research is a study case in Muhammadiyah University of Purwokerto. Even so, it does not cover the possibility to apply in other locations with the same case. Also it can be developed to determine the correlation or the effect on performance of lecturer competence in carrying out its function as educators.

**Acknowledgments**

Authors thank to DP2M - The Ministry of Education and Culture - Directorate General of Higher Education (DIKTI) through KOPERTIS Region VI who has provided funds in the implementation of this research.

## References
[1] D.E. Surya, "The Lecturers Competence of Standardization Services to Students", Scientific Magazine Unikom, Vol. 6 No. 2, 2011, pp. 157 – 168.
[2] Trianto, *et al*. The Rights and Obligations of the Juridical Review of Educator According to the ACT on Teachers and Lecturers, Jakarta: Prestasi Pustaka, 2006
[3] E. Ardianto, Public Relations a Practical Approach, Tips Become Communicators in Their Dealings with The Public and The Community, Bandung: Pustaka Bani Quraisy, 2004.
[4] The Legislation of the Teachers and Lecturers of Republic Indonesia Number 14, 2005
[5] S. Kusumadewi, "Fuzzy Quantification Theory I to Analysis the Relationship Between Performance Assessment of Lecturers by Students, the Attendance of Lecturers, and Students Graduation", Scientific Journal Media Informatika, Vol. 2, No. 1, 2004, pp. 1-10.
[6] S. Kusumadewi, and H. Purnomo, Application of Fuzzy Logic for Decision Support, Yogyakarta: Graha Ilmu, 2004.

**Hindayati Mustafidah** Master's degree in computer science at Gadjah Mada University (UGM) graduating with honors in 2002, Lecturer at the Muhammadiyah University of Purwokerto in Informatics Engineering. Several times as presenter in national and international seminars; more than 20 papers has been published in national journals and international digital libraries (IEEE *Xplore* Digital Library) and has published two books; the interested area of research is the field of intelligent systems; and incorporated in the profession organization of Indonesian Computer Electronics and Instrumentation Support Society (IndoCEISS).

**Suwarsito** Master of fisheries at Bogor Agricultural University (IPB); Lecturer at the Muhammadiyah University of Purwokerto in Geography Education; the interested area of research are educational research fields and the science of waters.